\documentclass[pdftex,twocolumn,epjc3]{svjour3}      
\usepackage{graphicx}
\usepackage{amssymb,mathtools,float,multirow}
\usepackage{amsmath,siunitx}
\usepackage{hyperref,multirow}
\hypersetup{colorlinks=true, citecolor=red, linkcolor=blue}
\usepackage{comment,color}
\newcommand{\revise}[1]{{\leavevmode\color{black}#1}}
\newcommand{\mrevise}[1]{{\leavevmode\color{black}#1}}

\begin{document}

\title{The flavor composition of astrophysical neutrinos after 8 years of IceCube: 
an indication of neutron decay scenario?}
\author{Andrea Palladino \thanksref{e0}}
\institute{DESY, Platanenallee 6, 15738 Zeuthen, Germany}
\thankstext{e0}{e-mail:andrea.palladino@desy.de}

\maketitle

\tableofcontents

\begin{abstract}
In this work we present an updated study of the flavor composition suggested by astrophysical neutrinos observed by IceCube. 
The main novelties compared to previous studies are the following: \textit{1)} we use the most recent measurements, namely 8 years of throughgoing muons and 7.5 years of High Energy Starting Events (HESE); \textit{2)} we consider a broken power law spectrum, in order to be consistent with the observations between 30 TeV and few PeV; \textit{3)} we use the throughgoing muon flux to predict the number of astrophysical HESE tracks.
We show that accounting for the three previous elements, the result favors surprisingly the hypothesis of neutrinos produced by neutron decay, disfavoring the standard picture of neutrinos from pion decay at 2.0$\sigma$ and the damped muons regime at $2.6 \sigma$, once the atmospheric background is considered. Although the conventional scenario is not yet completely ruled out in the statistically and alternative interpretations are also plausible, such as an energy spectrum characterized by a non trivial shape, this intriguing result may suggest new directions for both theoretical interpretation and experimental search strategies.
\end{abstract}

\maketitle

\section{Introduction}
In 2012 a diffuse flux of high energy neutrinos has been discovered by IceCube, a neutrino telescope placed in the South Pole \cite{Aartsen:2013jdh}. Since then, several theoretical works have been written to interpret the flavor composition observed by IceCube \cite{Vissani:2013iga,Mena:2014sja,Palladino:2015zua,Boucenna:2015tra,Bustamante:2015waa,Vincent:2016nut,Sui:2018bbh,Bustamante:2019sdb}. In the first few years after the detection, it was emphasized the inconsistency between the expected background for tracks contained in the High Energy Starting Events (HESE) dataset and the number of detected tracks \cite{Mena:2014sja}. That paper points out a \lq\lq misunderstanding of the expected background events or even more compellingly, some exotic physics which deviates from the standard scenario\rq\rq. Particularly when all the HESE are considered, as in \cite{Mena:2014sja}, the main issue is that the number of expected atmospheric tracks is larger that the number of detected tracks, favoring a null track to shower ratio for astrophysical neutrinos, with no known astrophysical interpretation so far. 
This issue was present in the three years dataset \cite{Aartsen:2014gkd} and it lingers into the 6 years dataset \cite{Aartsen:2017mau}. In \cite{Aartsen:2015ivb} the study of the flavor composition has been conducted using only events having deposited energy larger than 60 TeV (where the background is expected to be negligible), showing that the usage of this subset of data restores the compatibility between the detected flavor composition and the one expected from astrophysical production mechanisms.

In order to avoid possible problems related to the atmospheric muon background that affects HESE tracks, in \cite{Palladino:2015zua} a new method was used, with the inclusion of the flux measured with throughgoing muons (TGM). The main reason is that the throughgoing muon flux \cite{Aartsen:2015rwa} represents the cleanest way to observe the flux of astrophysical muon neutrinos, due to the high energy threshold (200 TeV) and to the absence of atmospheric muons, since they have no possibility to cross the Earth and reach the detector placed in the opposite hemisphere. On the contrary HESE tracks are likely to be largely affected by atmospheric muons, as remarked in \cite{Aartsen:2014gkd,Aartsen:2017mau}. \mrevise{Particularly the background expected for HESE tracks is larger than the number of the detected ones.} 
\mrevise{On the contrary, the flux measured using throughgoing muons is: \textit{i)} free from the contamination of atmospheric muons, since they are stopped inside the Earth; \textit{ii)} in a negligible manner contaminated by conventional atmospheric neutrinos, due to the high energy threshold of 200 TeV; \textit{iii)} marginally contaminated by prompt neutrinos, at level of 20\%, as can be estimated using the signalness contained in Tab.4 of \cite{Aartsen:2016xlq}.}
However HESE are also very important, since thanks to them the diffuse flux of astrophysical neutrinos has been measured for the first time \cite{Aartsen:2013jdh}. Moreover they are sensible to all neutrino flavors and the atmospheric background that affects showers is relatively low, due to the innovative veto technology. 

\mrevise{Following the previous discussion, we formulate} the main hypothesis of this work:

\begin{quote}
\mrevise{the expected number of astrophsyical HESE tracks is computed according to the shape and the normalization suggested by the throughgoing muon flux above 200 TeV and following the shape suggested by HESE below this energy. This is what we call \lq\lq baseline model\rq\rq \ in the rest of the work.
The number of astrophysical showers, instead, is extracted from the HESE dataset, accounting for the atmospheric background (that is however small for showers).} 
\end{quote}

Since this method does not rely on any assumption on the background that affects HESE tracks, it is the cleanest way to compute the observed track to shower ratio of astrophysical neutrinos and to compare it with the expected theoretical ones. We also demonstrate that the way in which we extrapolate the throughgoing muon flux below 200 TeV affects only marginally (at level of 10\%) the expected number of astrophysical HESE tracks. \mrevise{In other words, the computation of the expected number of HESE tracks depends mostly on the muon neutrino flux already measured.}




The current public available data consist of 7.5 years of High Energy Starting Events (HESE) and 8 years of throughgoing muons, therefore the analysis of the flavor composition can be much more powerful and accurate compared to the past.
Motivated by this argument, in this work we re-analyze the flavor composition observed by IceCube taking into account all the most recent measurements and using a spectrum that can describe the data between $\sim 30$ TeV and few PeV. 
Before the IceCube measurements, the flavor composition was already considered a powerful tool to understand the origin of high energy neutrinos \cite{Lipari:2007su}, that remains still a mystery.  Indeed, up to now, only one neutrino has been associated with an identified object \cite{IceCube:2018dnn}, but all the other neutrinos remain without any confirmed counterpart. 

The work is structured as follows. In Sec.\ref{sec:method} we discuss the method used to compute the flavor composition at Earth, how to convert it into an observable quantity and how to get information on the observed track to shower ratio. In Sec.\ref{sec:res} we present the result and we discuss the implications in Sec.\ref{sec:disc}. We conclude the work with Sec.\ref{sec:conc}.

\begin{figure*}[t]
\centering
\includegraphics[scale=0.45]{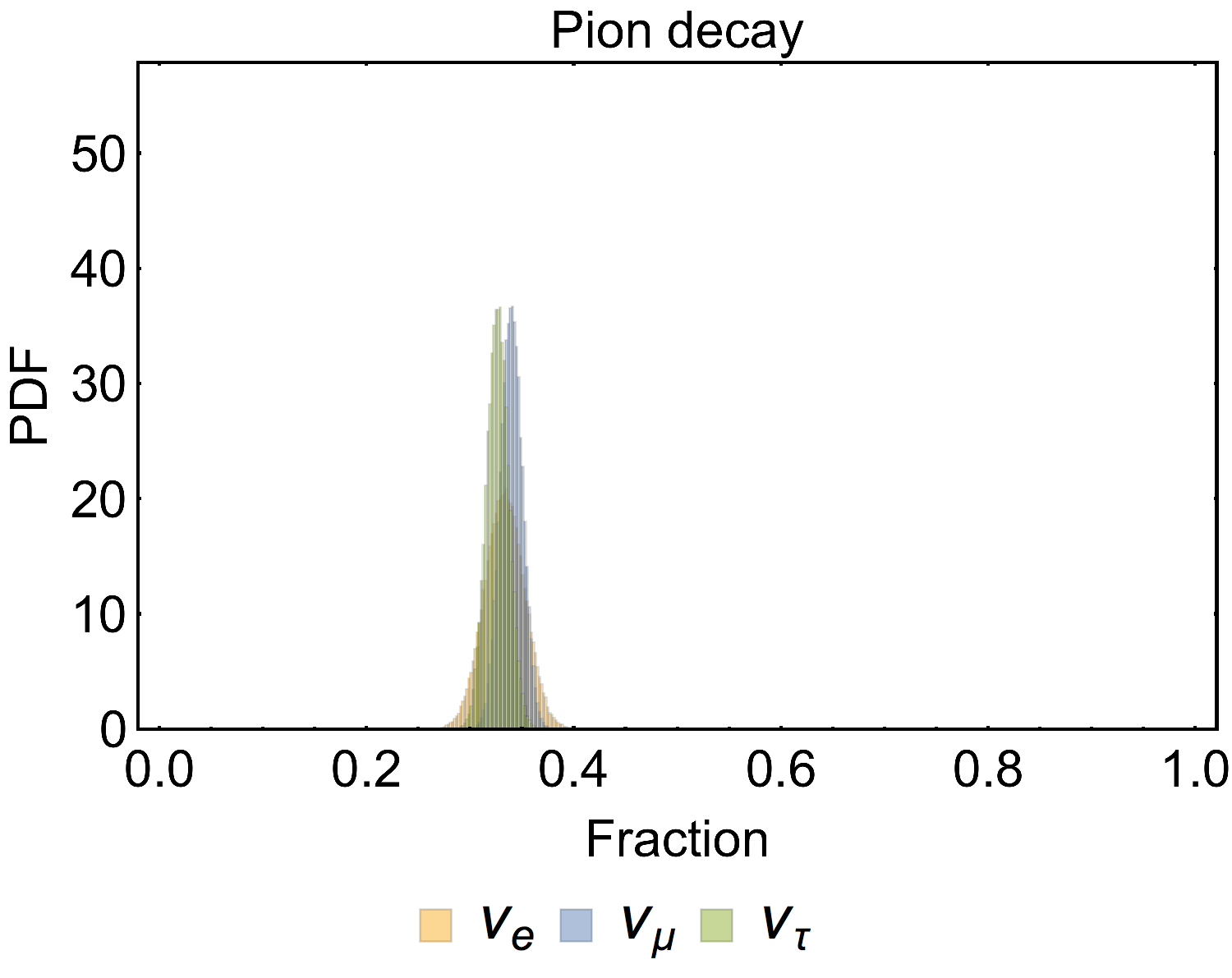}
\includegraphics[scale=0.45]{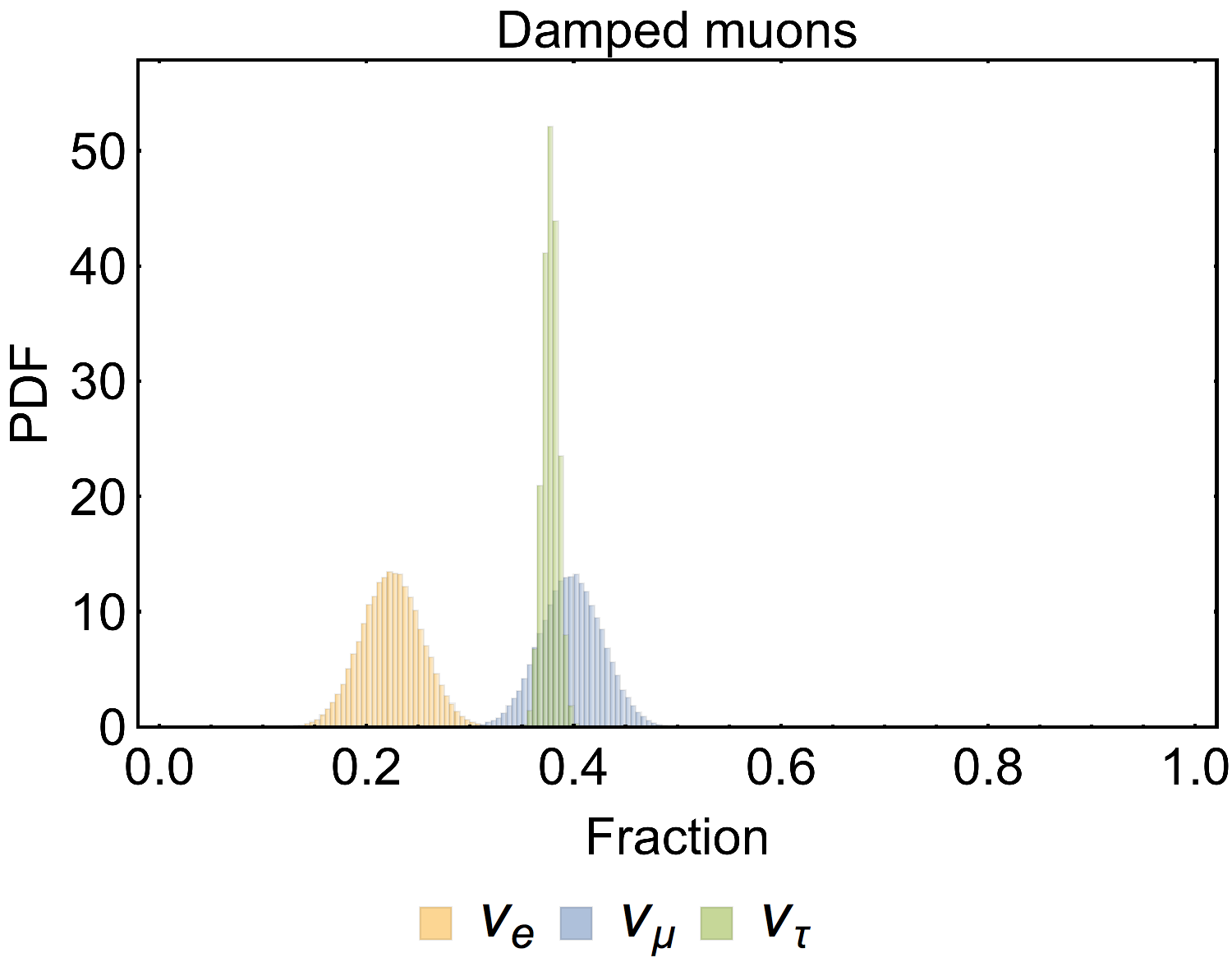}
\includegraphics[scale=0.45]{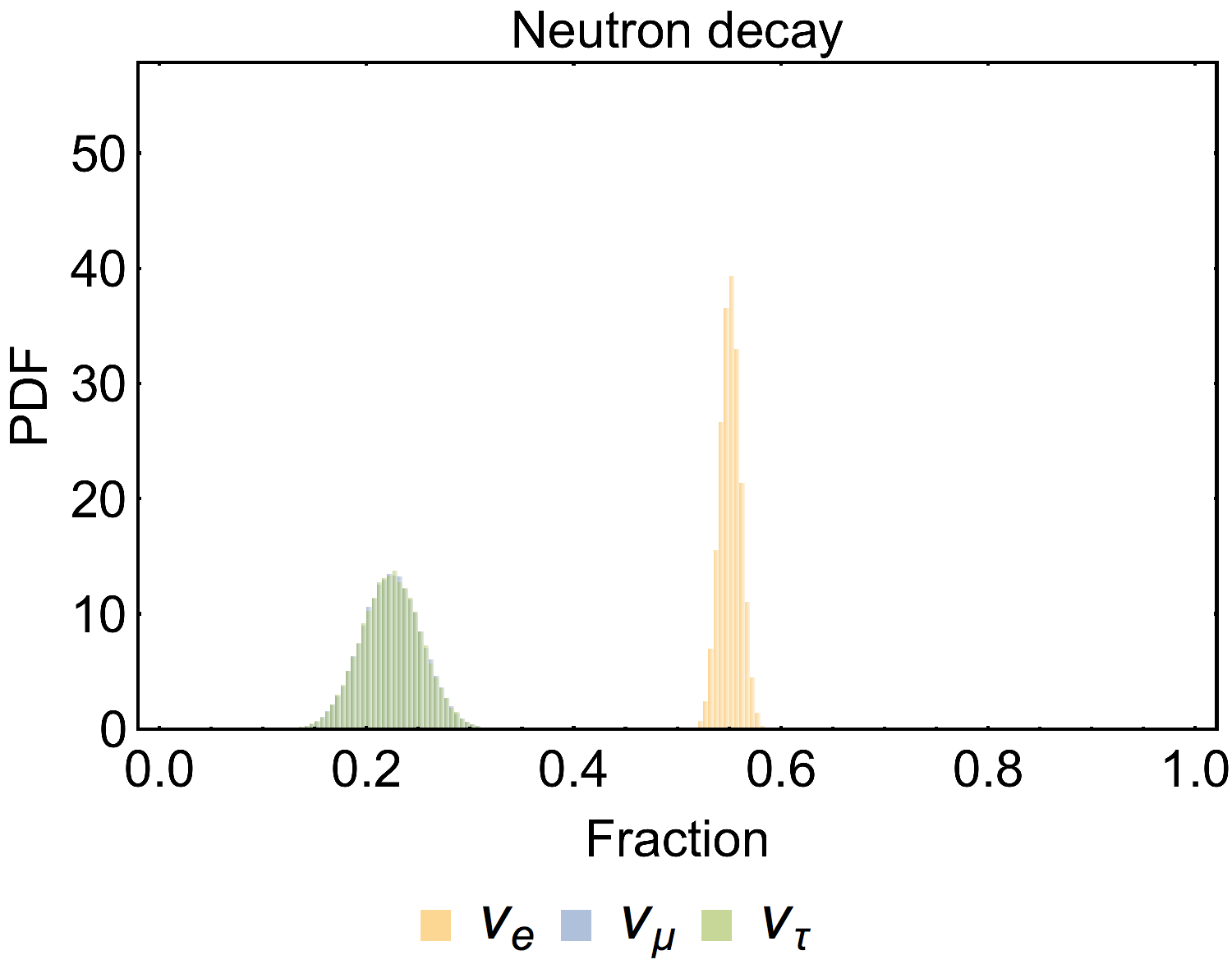}
\caption{Flavor composition of astrophysical neutrinos expected at Earth after neutrino oscillations, assuming that they are produced by pion decay (left panel), pion decay with damped muons (middle panel) and neutron decay (right panel). In the last case there is overlap between the fraction of $\nu_\mu$ and $\nu_\tau$ at Earth and the two distributions are not distinguishable in the figure.}
\label{fig:flavor}
\end{figure*}

\section{Method}
\label{sec:method}
\subsection{The theoretical flavor composition}
The propagation of cosmic TeV-PeV neutrinos \cite{Bilenky:1978nj} includes a long oscillation phase,
order of $\sim 10^{13}$ for neutrinos of 100 TeV coming from a distance of 1 Gpc. As a consequence only the average values of the oscillation probability are astrophysical observables.
In this work we compute the oscillations of astrophysical neutrinos using the \lq\lq natural parametrization\rq\rq \ introduced in \cite{Palladino:2015vna}. It permits to compute easily the uncertainties associated to the oscillation probabilities, using three Gaussian parameters $P_0$, $P_1$, $P_2$, where $P_0 \gg P_1 \simeq P_2$. 
\revise{The expressions of these  
parameters in terms of the conventional oscillation parameters 
(3 mixing angles and one CP violating phase) are:
\begin{align}
P_0 &=\frac{1}{2} \left[(1-\epsilon)^2 \left( 1-\frac{\sin^2 (2\theta_{12})}{2}\right)+\epsilon^2 -\frac{1}{3} \right] \label{eq:p0}\\[2mm]
P_1 &=\frac{1-\epsilon}{2}\left(\gamma\cos2\theta_{12}+\beta\frac{1-3\epsilon}{2} \right)\label{eq:p1}\\[2mm]
P_2 &=\frac{1}{2}\left[\gamma^2 + \frac{3}{4}\beta^2 (1-\epsilon)^2 \right] \label{eq:p2}
\end{align}
where
\begin{align*}
\epsilon &= \sin^2\theta_{13} \qquad \alpha = \sin\theta_{13}\cos\delta\sin2\theta_{12}\sin2\theta_{23} \\[1mm] \beta &= \cos2\theta_{23} \qquad \gamma = \alpha - \frac{\beta}{2}\cos2\theta_{12}(1+\epsilon)\end{align*}
Therefore $P_0$, $P_1$ and $P_2$ are not single values but distributions. Even if the distributions are not gaussian, we will use the gaussian approach that it is sufficiently accurate, as explained in \cite{Palladino:2015vna}.
}

Following the natural parametrization the oscillation probabilities $P_{\ell\ell^{'}}$ (where $\ell$ and $\ell^{'}$ denotes the initial and the final neutrino flavor) are given by the elements of the following matrix:
\begin{equation} \label{pupp}
\mathcal{P}=\left( \begin{array}{ccc}
\frac{1}{3}+2 P_0 & \frac{1}{3}-P_0+P_1 & \frac{1}{3}-P_0-P_1 \\
 & \frac{1}{3}+\frac{P_0}{2}-P_1+P_2 & \frac{1}{3}+\frac{P_0}{2}-P_2 \\
 &  & \frac{1}{3}+\frac{P_0}{2}+P_1+P_2 \end{array} \right)   \nonumber
 \end{equation} 
The matrix acts on the vector containing the flavor composition before oscillations   $\xi^0=(\xi_e^0,\xi_\mu^0,\xi_\tau^0)$ just as $\xi=\mathcal{P}\ \xi^0$, giving 
the vector of fluxes observed after oscillations, $\xi=(\xi_e,\xi_\mu,\xi_\tau)$. 
The \revise{average values and the uncertainties} of the natural parameters are taken from \cite{Palladino:2015vna} (based on the knowledge of neutrino oscillations given in \cite{Gonzalez-Garcia:2014bfa}) and they are equal to:
\begin{eqnarray}
P_0  &=& 0.109 \pm 0.005  \nonumber \\
P_1 &=& 0.000 \pm 0.029  \nonumber \\
P_2 &=& 0.010 \pm 0.007  \nonumber
\end{eqnarray}

Concerning the initial flavor composition we assume three conventional astrophysical scenarios:
\begin{itemize}
\item neutrinos are produced via charged pion decay, following $\pi^+ \rightarrow \mu^+ \nu_\mu \rightarrow e^+ \nu_\mu \bar{\nu}_\mu  \nu_e$ or $\pi^- \rightarrow \mu^- \bar{\nu}_\mu \rightarrow e^- \bar{\nu}_\mu \nu_\mu  \bar{\nu}_e$. In this process the flavor composition at the source is equal to $(\xi_e^0:\xi_\mu^0:\xi_\tau^0)=(1:2:0)$. We do not distinguish between neutrinos and antineutrinos in this work, \revise{since the only channel to observe astrophysical antineutrinos is the Glashow resonance (see Sec. \ref{sec:glashow}) and these events are still not observed in the present neutrino telescope. Therefore current observations are only sensitive to the sum of neutrino and antineutrino fluxes.}
\item neutrinos are produced by pion decay in astrophysical environment with strong magnetic fields (\revise{$\sim 10^5--10^6$ Gauss}) \cite{Hummer:2010ai}. Under this assumption, muons lose a significant fraction of energy before decaying, therefore high energy neutrinos are only created by the first part of the previous chain, i.e.\   $\pi^+ \rightarrow \mu^+ \nu_\mu$ or $\pi^- \rightarrow \mu^- \bar{\nu}_\mu$. In this case the initial flavor composition is equal to $(\xi_e^0:\xi_\mu^0:\xi_\tau^0)=(0:1:0)$;
\item neutrinos are created by the decay of neutrons, according to the process $n \rightarrow p \ e^- \bar{\nu}_e$. In this scenario the initial flavor composition is equal to $(\xi_e^0:\xi_\mu^0:\xi_\tau^0)=(1:0:0)$.
\end{itemize}

Using the matrix $\mathcal{P}$ defined above we can compute the neutrino oscillations, obtaining the flavor composition at Earth as $\xi=\mathcal{P} \xi_0$. The flavor compositions obtained at Earth are reported in Tab.\ref{tab:flavor} and Fig.\ref{fig:flavor}. We notice that the uncertainties on the final flavor composition are not the same for all production mechanisms; this is related to the fact that the knowledge of the natural parameters is not equally good, since $\Delta P_1 \gg \Delta P_0 \simeq \Delta P_2$.

\begin{table}[t]
\caption{Flavor composition expected at Earth for the three different production mechanisms, accounting for the uncertainties on the neutrino oscillations.}
\begin{center}
\begin{tabular}{cccc}
\hline
& $\xi_e$ & $\xi_\mu$ & $\xi_\tau$ \\ 
\hline
$\pi$ decay & $0.33 \pm 0.02$ & $0.34 \pm 0.01$ & $0.33 \pm 0.01$ \\ 
damped  $\mu$ & $0.22 \pm 0.03$ & $0.40 \pm 0.03 $ & $0.38 \pm 0.01$ \\ 
$n$ decay & $0.55 \pm 0.01$ & $0.225 \pm 0.03$ & $0.225 \pm 0.03$ \\ 
\hline
\end{tabular}
\end{center}
\label{tab:flavor}
\end{table}%

\subsection{The theoretical track to shower ratio expected from astrophysical scenarios}
\label{sec:glashow}
The flavor composition is not a direct observable in IceCube, since only two types of event topologies are so far identified in the modern neutrino telescopes, namely tracks and showers \cite{Aartsen:2013vja}. Neutrinos are generally detected thanks to the deep inelastic scattering \cite{Gandhi:1998ri}, looking at the secondary particles produced after the interaction between neutrinos and nucleons. Tracks are produced by the interaction of $\nu_\mu$ via charged current interaction, while showers are produced by all the other processes, i.e.\ charged current interactions of $\nu_e$ and $\nu_\tau$ and neutral current interactions of whatever neutrino. In principle there are two other processes that permit to identify $\bar{\nu}_e$ and $\nu_\tau$: the Glashow resonance \cite{Glashow:1960zz} and the double cascades \cite{Learned:1994wg}, but they are still not observed.\footnote{Up to now there are no publications concerning the detection of this kind of events, although 1 candidate resonant event has been presented at the conference UHECR 2018 and 2 candidate $\nu_\tau$ have been presented at the conference NEUTRINO 2018.}

\revise{The first analysis of the flavor composition observed by IceCube has been presented for the first time in \cite{Mena:2014sja}. However, }
from the previous discussion, it follows that the observable quantity is not directly the flavor composition but the ratio between the number of tracks and the number of showers, abbreviated \lq\lq track to shower ratio\rq\rq \ in the following of this work. The analysis of the track to shower ratio has been \revise{adopted} in \cite{Palladino:2015zua} and in this work we update it, using the most recent IceCube measurements after about 8 years of exposure. In order to do that we need to include information on the incident astrophysical neutrino spectrum and on the response function of the detector.

\begin{figure*}[t]
\centering
\includegraphics[scale=0.66]{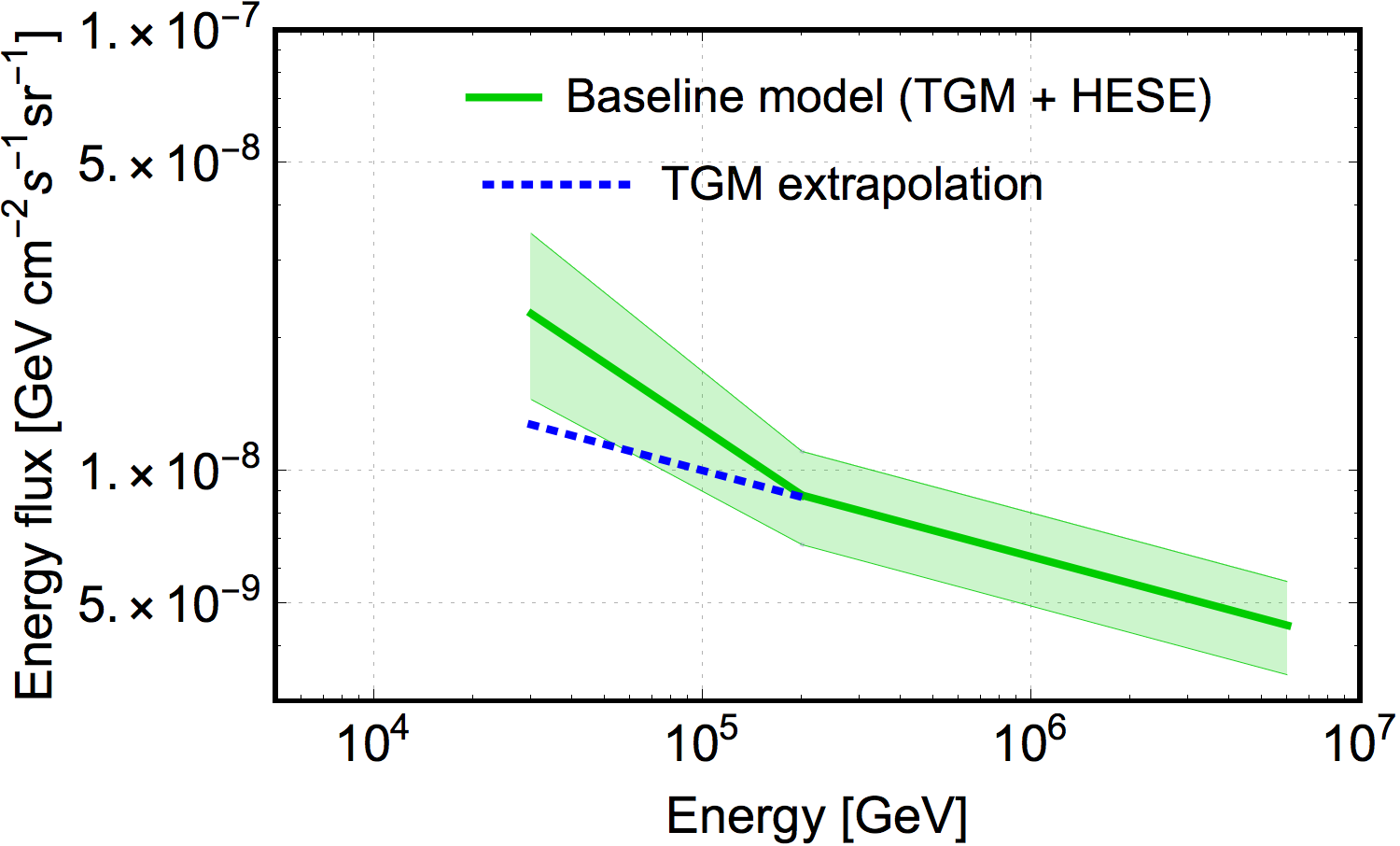}
\includegraphics[scale=0.615]{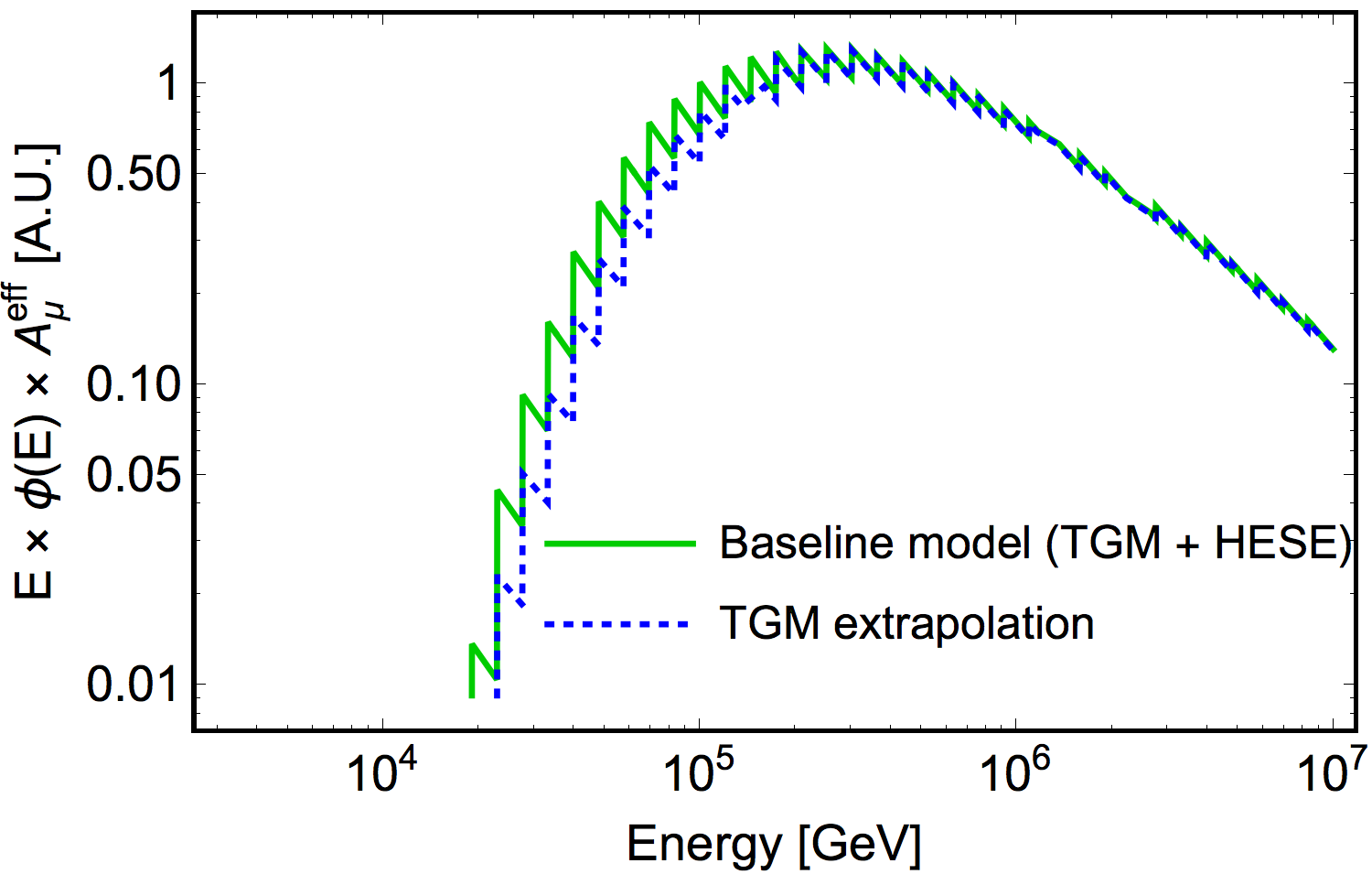}
\caption{On the left panel: the baseline single flavor spectrum of astrophysical neutrinos is represented with \revise{a solid green curve.} It is obtained following the shape and the normalization suggested by throughgoing muons above 200 TeV \cite{Aartsen:2017mau} and shape suggested by HESE below 200 TeV \cite{Aartsen:2015knd}. \revise{The green band is related to the uncertainty on the measured normalization at 100 TeV ($\sim 30\%$ for both HESE and TGM).} The blu dotted line denotes the extrapolation to lower energies following the throughgoing muon shape. On the right panel we represent the product between flux and muon neutrino effective area, in arbitrary units.}
\label{fig:icedata}
\end{figure*}

\paragraph{Spectrum:} Up to now there are measurements of the astrophysical neutrino spectrum covering different energy ranges and sky locations. 
Throughgoing muons, only sensible to $\nu_\mu$ from Northern sky above 200 TeV, suggest a hard spectrum $\propto E^{-2.2 \pm 0.1}$ \cite{Aartsen:2017mau}. On the other hand High Energy Starting Events (HESE), that are sensitive to the all flavor flux from both hemispheres, suggest a softer spectrum between $\propto E^{-2.5 \pm 0.1}$ \cite{Aartsen:2015knd} between 30 TeV \revise{and 3 PeV.\footnote{A recent analysis presented at ICRC 2017 \cite{Aartsen:2017mau} shows even a softer spectrum for HESE, according to $\propto E^{-2.9 \pm 0.3}$. On the other hand this result is in contrast with the indication coming from the cascade dataset extending lower to $\sim$ TeV energy, that shows a $\propto E^{-2.44 \pm 0.08}$ \cite{Aartsen:2017mau}. For this reason we continue to use the information provided in \cite{Aartsen:2015knd} for the HESE spectrum.} Moreover let us notice that about 90\% of HESE have an energy smaller than 200 TeV while all the throughgoing muons have an energy larger than 200 TeV}. Therefore \revise{it is reasonable} having trust of the spectral shape suggested by throughgoing muons above 200 TeV and of the spectral shape suggested by HESE below 200 TeV. This is our baseline choice for the spectrum of astrophysical neutrinos $\phi(E)$ and it is represented in the left panel of Fig.\ref{fig:icedata}. The normalization above 200 TeV replicates the normalization of the throughgoing muon spectrum given in \cite{Aartsen:2017mau}. Let us remark that due to neutrino oscillations and standard astrophysical mechanisms, we expect the same spectral shape for all flavors at Earth. Only the normalization can change according to the production mechanism, as shown in Tab.\ref{tab:flavor}. The idea of a two component spectrum is plausible and it has been already discussed in several theoretical works \cite{Chen:2014gxa,Palomares-Ruiz:2015mka,Palladino:2016zoe,Palladino:2016xsy,Palladino:2018evm}.

\paragraph{Response of the detector}: in order to convert the theoretical flavor composition in the observable track to shower ratio, we also need to know the response of the detector to neutrinos. This information is contained in the angle-averaged effective areas $A_\ell^{\text{eff}}$, provided by IceCube for each neutrino flavor $\nu_\ell$ \cite{Aartsen:2013jdh}. Using all the information discussed above, we compute \revise{the parameters $C_i$, that denote in which way the flavor composition is modified by the detector. The parameters $C_i$ will be used in the computation of the track-to-shower ratio and they are defined as follows:}
\begin{eqnarray}
C_e &=& C_0 \int_0^\infty dE \ A_e^{\text{eff}} \phi(E)  \nonumber \\
C_\mu^t &=& C_0 \ \eta \int_0^\infty dE \ A_\mu^{\text{eff}} \phi(E)  \nonumber \\
C_\mu^s &=& C_0 \ (1-\eta) \int_0^\infty dE \ A_\mu^{\text{eff}} \phi(E)  \nonumber \\
C_\tau &=& C_0 \int_0^\infty dE \ A_\tau^{\text{eff}} \phi(E) \nonumber 
\end{eqnarray}
where $C_0=(\sum_{\ell=e,\mu,\tau} \int_0^\infty dE \ A_\ell^{\text{eff}} \phi(E))^{-1}$ and $\eta=0.8$ as in \cite{Palladino:2015zua}, denoting the fraction of the muon neutrino effective area that is connected to charged current interactions\footnote{A common mistake consists in believing that $\sim 18\%$ of tracks are produced by $\nu_\tau$. Considering that muons from tau decay take only $\sim 1/3$ of the tau energy, the contribution of $\nu_\tau$ to the tracks is only 2\% for an $E^{-2}$ spectrum and it decreases for softer spectra.}.

The apex $t$ or $s$ denotes the topology of the event, namely track or shower.
The values of these parameters, obtained for our baseline spectrum $\phi(E)$, are equal to $C_e=0.49, C_\mu^t=0.17, C_\mu^s=0.04, C_\tau=0.30$. 
\revise{Let us notice that using an ideal detector that does not modify the flavor composition, we would obtain $C_e = C_\tau = C_\mu^s+C_\mu^t$; this is not true in reality due to the different energy deposited by neutrinos having different flavors.}


The way to convert the flavor composition expected at Earth in the track to shower ratio $r$ is given by the following expression, using the previous parameters $C_\ell$:
\begin{equation}
r^{\text{th}}(\xi_e,\xi_\mu)= \frac{\xi_\mu C_\mu^t}{\xi_e C_e + \xi_\mu C_\mu^s + \xi_\tau C_\tau}
\label{eq:rtheory}
\end{equation}
Let us recall that $\xi_e+\xi_\mu+\xi_\tau=1$, therefore there are only 2 independent variables. The theoretical track to shower ratios $r_{\text{th}}$ obtained for the three different production mechanisms \revise{using the baseline spectrum} are shown in \revise{the right panel of Fig.\ref{fig:trtosh}; namely pion decay (orange bars), damped muons (red bars) and neutron decay (green bars)}. They are equal to $0.21 \pm 0.01$ for the pion decay scenario, to $0.29 \pm 0.04$ for the damped muons scenario and $0.11 \pm 0.02$ for the neutron decay scenarios. Assuming $(\xi_e:\xi_\mu:\xi_\tau)=(1:1:1)$ at Earth, we obtain a track to shower ratio equal to 0.21 using our baseline spectrum. \revise{The flavor composition is \textit{always} assumed to be energy-independent in the following of the work.}

\mrevise{Before proceeding a clarification is necessary. The spectrum suggested by HESE $\propto E^{-2.5}$  reflects the behavior of the measurements between 30 TeV and few PeV. On the other hand in our baseline model we are only using this shape for $E < 200 \mbox{ TeV}$. Limiting the HESE data to the energy between 30 TeV and 200 TeV would result in a different spectral shape, suggesting probably a softer spectrum, since the HESE above 200 TeV are in agreement with the throughgoing muons measurements (i.e.\ with a hard spectrum). 
However the analysis of the 4 years shower dataset above 1 TeV, presented in Sec.3 of \cite{Aartsen:2017mau}, suggest an $\propto E^{-2.48 \pm 0.08}$ spectrum at lower energies. In conclusions, there are no valid reasons to use a spectrum softer than $E^{-2.5}$ below 200 TeV.}

\subsection{The track to shower ratio of astrophysical neutrinos in IceCube}
\label{sec:trtosh}
The expected track to shower ratio computed in the previous section has to be compared to the detected one. In order to do that we consider the most recent HESE data, presented in \cite{taboada2018}. This dataset consists of 113 events, including 30 tracks, 81 showers and 2 not classified events (that were already present in the previous datasets), detected after 7.5 years of exposure.

The computation to predict the observed track to shower ratio is complicated, as we need to appropriately include all sources of background.
Let us notice that both in \cite{Aartsen:2014gkd,Aartsen:2017mau} the expected atmospheric background for HESE tracks is larger than the observed number of tracks, when all HESE are considered. This represents an issue for the computation of the track to shower ratio, since it would indicate that no astrophysical tracks are present in the HESE sample. This information was used in \cite{Mena:2014sja} to claim a possible tension between neutrino oscillations and IceCube measurement. On the other hand the IceCube analysis, performed using only events above 60 TeV (where the atmospheric background is expected to be negligible), claims an opposite result compared to \cite{Mena:2014sja}, showing that the observed flavor composition is in agreement with the damped muon scenario and compatible with the pion decay \cite{Aartsen:2015ivb}. However even in this case a problem remains: a large number of tracks (more than 20 tracks in the 6 years dataset \cite{Aartsen:2017mau}) is expected between 30 TeV and 60 TeV but not detected. Both these analyses depend on the assumed background for HESE tracks in the considered energy region and they give completely different results, since they use two different energy thresholds. Let us clarify that the main source of background is represented by atmospheric muons in this case, not by atmospheric neutrinos.

In \cite{Palladino:2015zua} it has been proposed a new method, that does not require any assumption and any knowledge of the background related to HESE tracks. This method consists in the computation of the expected number of HESE tracks, using the well measured throughgoing muon flux and the muon neutrino effective area. Although this flux is only measured above 200 TeV, we demonstrate that the extrapolations to lower energies affects only marginally the expected number of HESE tracks. In other words, \emph{the most important part of the spectrum for this kind of calculation is the one already measured.}

\begin{figure*}[t]
\centering
\includegraphics[scale=0.63]{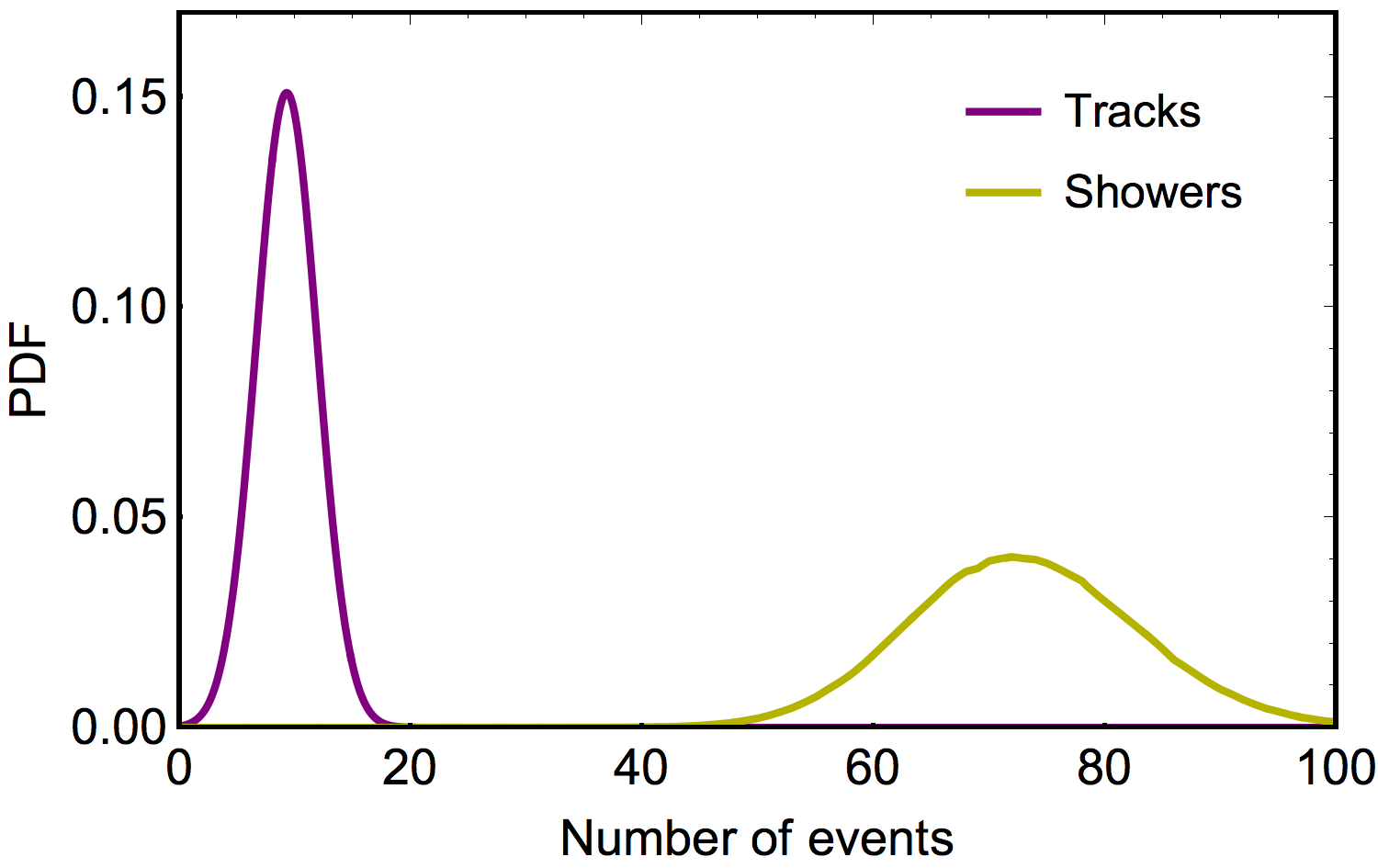}
\includegraphics[scale=0.6]{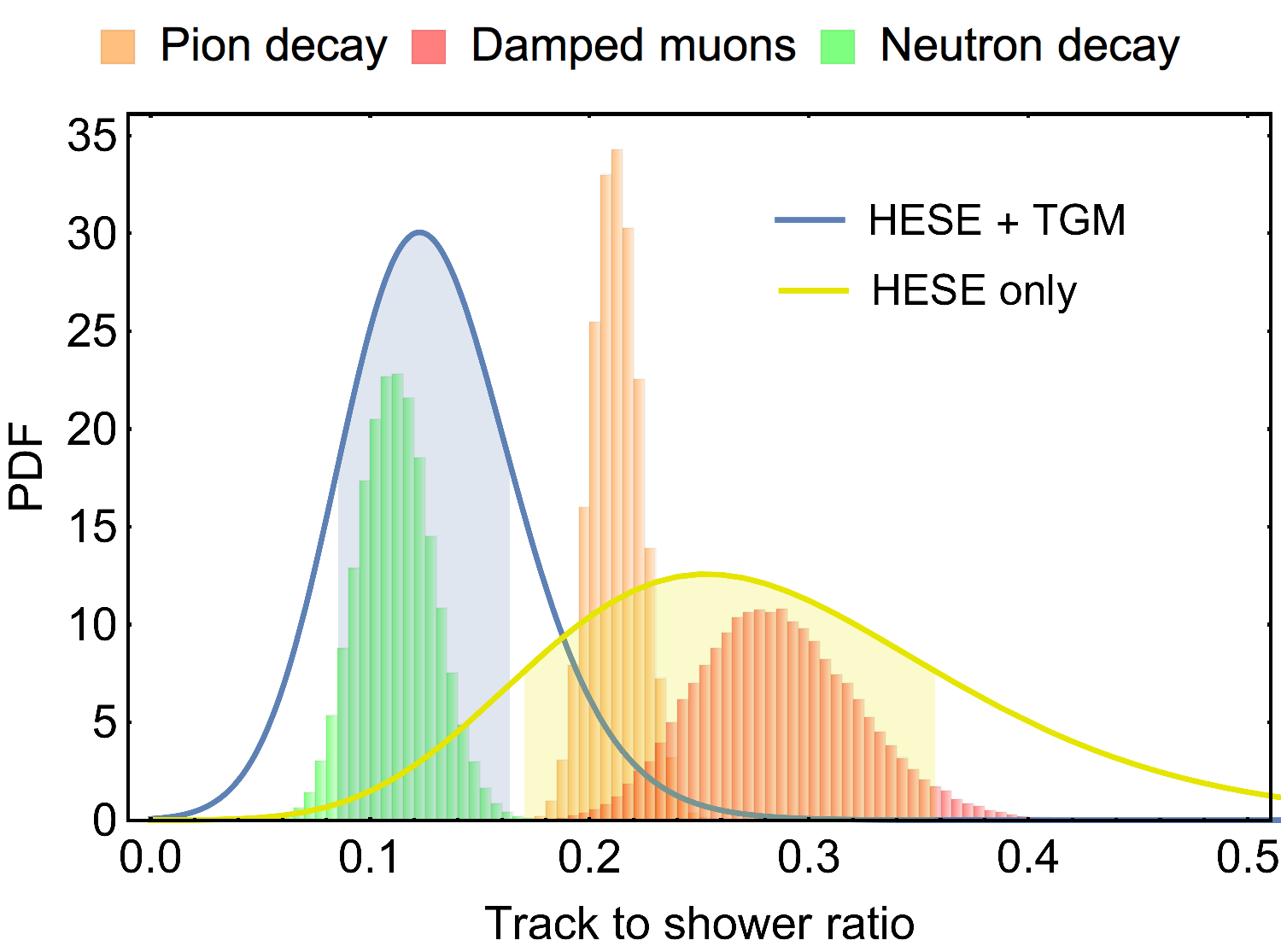}
\caption{On the left panel: likelihood for the number of astrophysical HESE tracks (purple curve) and astrophysical HESE showers (yellow curve) after 7.5 of exposure with IceCube. On the right panel: expected astrophysical track to shower ratio for different production mechanisms (orange bars for pion decay, red bars for damped muons, green bars for neutron decay) compared to the track to shower ratio derived by IceCube observations, using the throughgoing muon flux + HESE (blue curve) and HESE above 60 TeV only (yellow region). The shaded regions show the 1$\sigma$ interval.}
\label{fig:trtosh}
\end{figure*}

\paragraph{Astrophysical tracks:} using the muon effective area and our baseline spectrum, we can compute the expected number of HESE tracks as follows:
$$
N_t^{\text{astro}} = 4 \pi \mbox{T} \ \eta \int_0^\infty \phi(E) A_\mu^{\text{eff}} \ dE
$$
where T=7.5 years.\footnote{In order to check the correctness of our procedure we checked that we are able to obtain the total number of astrophysical events reported in Tab.4 of \cite{Aartsen:2014gkd}, within 4\% of accuracy. Using the $E^{-2}$ spectrum mentioned in that paper and the exposure of 2.7 years we obtain 23.67 events, while in the table is quoted 23.8. Using the $E^{-2.3}$ spectrum we obtain 22.8 events, versus the 23.7 events quoted by IceCube. The obtained track to shower ratio obtained by us is 0.22 for $E^{-2}$ and 0.21 for $E^{-2.3}$, while the values quoted by IceCube are 0.23 and 0.22 respectively. It confirms that our approach is adequate for the purpose.} Using the baseline spectrum represented in the left panel of Fig.\ref{fig:icedata} we obtain: 
\begin{equation}
N_t^{\text{astro}}= 9.3^{+2.6}_{-2.3} 
\label{eq:nastro}
\end{equation}
The asymmetric uncertainty is related to the asymmetric uncertainty on the normalization of the throughgoing muon flux \cite{Aartsen:2017mau}.
On the other hand if we extrapolate the throughgoing muon flux at lower energies following the $E^{-2.2}$ flux, we obtain $N_t^{\text{astro}}=8.4^{+2.1}_{-1.9}$, i.e.\ a discrepancy of 10\%. It means that the extrapolation of the spectrum has only a minor role in this calculation and it is confirmed by the right panel of Fig.\ref{fig:icedata}, in which the differential number of expected events is represented as a function of energy.  
\revise{Following our baseline model, the likelihood for the astrophysical tracks is then given by a function $\mathcal{L}_t(n_t)$ having a maximum in $n_t=9.3$ and having the integral equal to 0.68 in the interval $n_t=9.3^{+2.6}_{-2.3}$. We choose a function $\mathcal{L}_t(n_t)$ consisting of two pieces of not normalized Gaussian functions $\mathcal{G}(n_t,\mu,\sigma)$, with $n_t=9.3$ as splitting point. The functions are characterized by having the same mean $\mu=9.3$ and different standard deviation $\sigma$, namely $\sigma = 2.6$ for  $n_t\geq9.3$ and $\sigma=2.3$ for $n_t < 9.3$. Then we normalize the two pieces of Gaussian function in order to obtain a continuous function. The integral of the likelihood correctly replicate the $1\sigma$ interval found above $N_t^{\text{astro}}= 9.3^{+2.6}_{-2.3}$, as follows:
 }
$$
\frac{\int_{n_t^1}^{n_t^2} \mathcal{L}_t(n_t) \ dn_t}{\int_{0}^{\infty} \mathcal{L}_t(n_t) \ dn_t} = 0.68 
$$
where $n_t^1=7.0$ and $n_t^2=11.9$ are the extremes of the $1\sigma$ region of the expected astrophysical HESE tracks. The likelihood \revise{denoting the number of astrophysical tracks} is represented in the left panel of Fig.\ref{fig:trtosh} \revise{using a purple curve}.

\begin{figure}[t]
\centering
\includegraphics[scale=0.605]{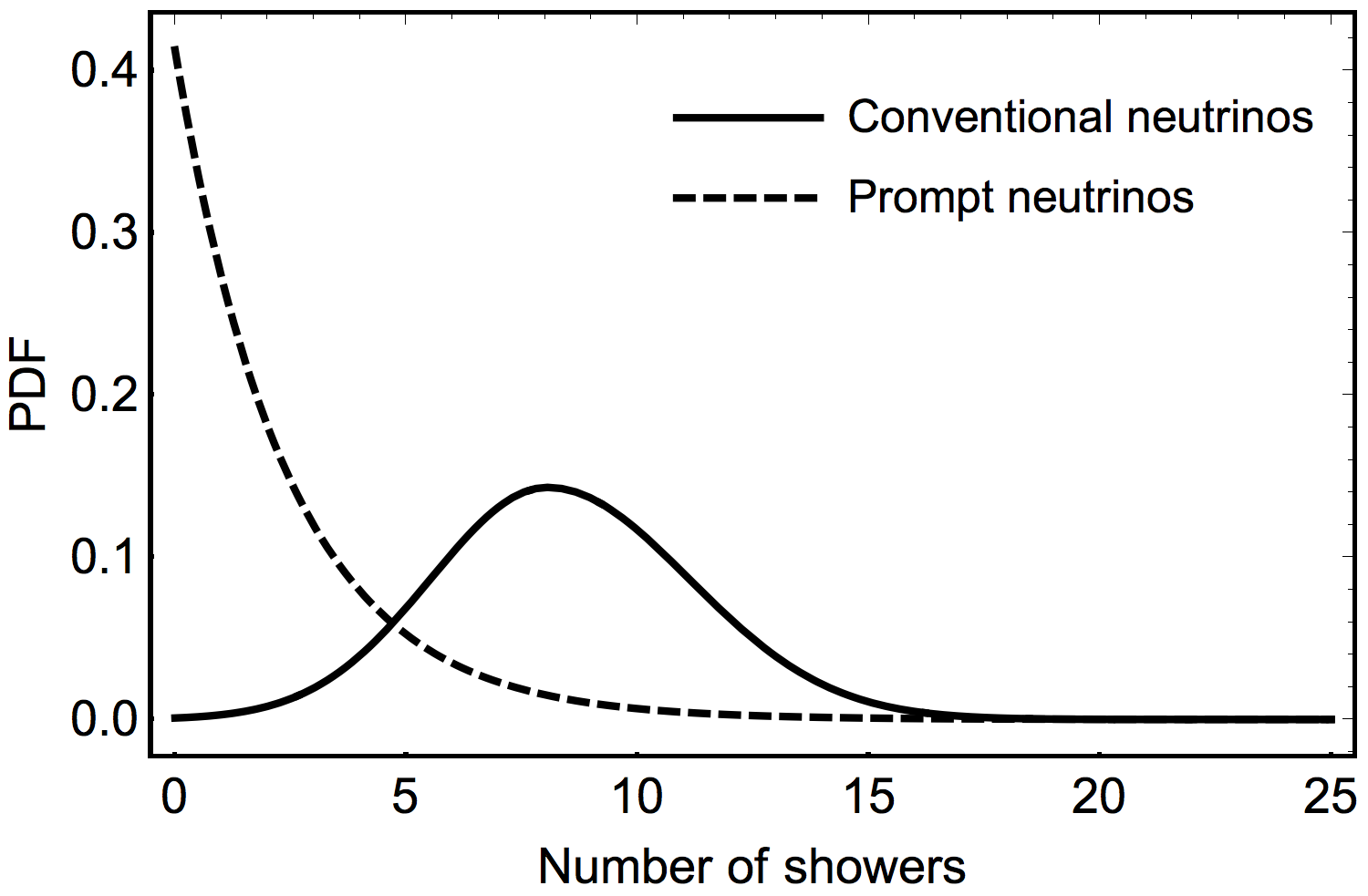}
\caption{The expected contribution of the atmospheric background to HESE showers in 7.5 years of exposure. The solid curve denotes the contribution expected from atmospheric neutrinos and atmospheric muons \cite{Aartsen:2014gkd}, while the dashed curve denotes the contribution expected from prompt neutrinos \cite{Aartsen:2014gkd,Aartsen:2016xlq}}
\label{fig:background}
\end{figure}

\paragraph{Astrophysical showers:} here we proceed to compute the number of astrophysical showers among the 81 showers contained in the 7.5 years HESE dataset. The showers are much less affected by atmospheric background compared to tracks; this is evident from Tab.4 of \cite{Aartsen:2014gkd}. In order to get the background expected in 7.5 years we scale in time the background of Tab.4  of \cite{Aartsen:2014gkd}, that refers to an exposure of 2.7 years. We obtain that after 7.5 the expected background from conventional neutrinos plus atmospheric muons\footnote{From Tab.4 of \cite{Aartsen:2014gkd} we read than \revise{approximately} 10\% of atmospheric muons can be identified as showers, probably due to the misidentification.} is equal to:
$$
N_{s}^{\text{conv}}= 8.0^{+3.0}_{-2.5}
$$
Following the same table the background associated to prompt neutrinos should be $N_{s}^{\text{prompt}} \leq 20$ at 90\% C.L. in 7.5 years. On the other hand that limit was derived based on \cite{Aartsen:2013eka}, in which the upper limit on prompt neutrinos was $3.8 \times \phi_{\text{ERS}}$, where $\phi_{\text{ERS}}$ is the theoretical flux of prompt neutrinos calculated in \cite{Enberg:2008te}. Recently the upper limit on prompt neutrinos has been improved, reaching the level of $1.06 \times \phi_{\text{ERS}}$ in \cite{Aartsen:2016xlq}. Therefore after 7.5 years we expect that prompt neutrinos give at the best fit a null contribution to HESE showers and they can contribute at level of:
$$
N_{s}^{\text{prompt}} < 5.6  \ \ \ \mbox{at 90\% C.L.}
$$
The likelihoods for the conventional background $\mathcal{L}_s^{\text{conv}}$ and for prompt neutrinos $\mathcal{L}_s^{\text{prompt}}$ are represented in Fig.\ref{fig:background}. \revise{For conventional atmospheric showers the likelihood is constructed in order to obtain the integral equal to 0.68 in the interval $8.0^{+3.0}_{-2.5}$; this function is constructed using two pieces of Gaussian functions, as explained below for the likelihood of astrophysical tracks. For atmospheric prompt neutrinos, instead, we consider an exponential function (being the experimental best fit equal to 0) under the assumption that the integral of the this likelihood is equal to 0.9 between 0 and 5.6}.

Now we have all the ingredient to compute the likelihood for the \revise{number of astrophysical showers $n_s$} that contribute to HESE showers, according to the following equations:
\begin{equation}
\begin{split}
\mathcal{L}_s(n_s) \propto \ \int_0^\infty dn_s^{\text{conv}} \int_0^\infty dn_s^{\text{prompt}} (n_s+n_s^{\text{conv}}+n_s^{\text{prompt}})^{N_s} \\
\exp[-(n_s+n_s^{\text{conv}}+n_s^{\text{prompt}})] \ \mathcal{L}_s^{\text{conv}}(n_s^{\text{conv}}) \ \mathcal{L}_s^{\text{prompt}}(n_s^{\text{prompt}})
\end{split}
\end{equation}
where  $N_s=81$ denotes the number of observed showers. The resulting number of astrophysical showers is equal to:
\begin{equation}
N_s^{\text{astro}}= 73.0^{+9.5}_{-10.2}
\label{eq:nshower}
\end{equation}
and the likelihood function $\mathcal{L}_s(n_s)$, \revise{denoting the number of astrophysical showers}, is represented in the left panel of Fig.\ref{fig:trtosh} \revise{using a yellow curve}.

The observed track to shower ratio can be computed using the following expression:
\begin{equation}
\mathcal{L}_r^{\text{obs}}(r) \propto \int_0^\infty \mathcal{L}_t(r \ n_s) \ n_s \ \mathcal{L}_s(n_s) \ dn_s
\label{eq:lik}
\end{equation}
after the changing of variable $n_t= r  n_s$. 

\section{Results}
\label{sec:res}
\paragraph{HESE + throughgoing muons:}
 the likelihood relative to the observed astrophysical track to shower ratio \revise{(see Eq.(\ref{eq:lik}))}, accounting for the throughgoing muon flux, is reported in the right panel of Fig.\ref{fig:trtosh} as a blue curve. The observed track to shower ratio $r_{\text{astro}}^{\text{HESE + TGM}}$ is equal to:
$$
r_{\text{astro}}^{\text{HESE + TGM}}  = 0.12 \pm 0.04
$$
\revise{and the shaded blue region of the right panel of Fig.\ref{fig:trtosh} denotes the $1\sigma$ interval.}

In order to define the compatibility between the observations and the theoretical expectations we use a statistical treatment, defining the function 
$$
D(\delta)^i  = \int_0^\infty \mathcal{L}^{\text{obs}}_r(r+\delta) \mathcal{L}_r^{\text{th},i}(r) dr
$$
where $\mathcal{L}^{\text{obs}}_r$ is the observed likelihood defined above and $ \mathcal{L}_r^{\text{th},i}$ is the theoretical track to shower ratio expected from  the production mechanism $i$. Then we calculate at how many $\sigma$ the value $\delta=0$, i.e.\ the null distance between these two distributions (\revise{i.e.\ $\mathcal{L}^{\text{obs}}$ and $\mathcal{L}_r^{\text{th},i}$}), is disfavored. In order to do that we cut the distribution \revise{D} in two points, at equal height, defining:
$$
\revise{\delta_1}= 0, \ \ \ \revise{\delta_2} \rightarrow D(\revise{\delta_2})=D(0)
$$
and we compute
$$
\mathcal{I}^i=\frac{\int_{\delta_1}^{\delta_2} D(\delta)^i \ d\delta}{\int_{-\infty}^{\infty} D(\delta)^i \ d\delta}
$$
\revise{After checking that the distributions $D^i$ are in good approximation normally distributed, we convert} the result of the previous integral in a number of $\sigma$, using a Gaussian approach (i.e.\ $0.68 \rightarrow 1\sigma, \ 0.95 \rightarrow 2\sigma, \ 0.997 \rightarrow 3\sigma$ etc...). We find that:
\begin{itemize}
\item the neutron decay scenario is the best option, resulting well compatible with the observed track to shower ratio;
\item the pion decay scenario is disfavored at 2$\sigma$;
\item the damped muon scenario is disfavored at $2.6\sigma$.
\end{itemize}

\paragraph{HESE only:} we also show the result that comes out from the conventional procedure, considering HESE above 60 TeV and \revise{accounting for} the background. In the 7.5 years HESE dataset we find 19 tracks and 51 showers above 60 TeV. 

Scaling the background reported in Tab.4 of \cite{Aartsen:2014gkd} with the exposure, the expected background consists of $\sim$ 6 tracks and $\sim$ 2 showers. Following the same procedure reported in Sec.\ref{sec:trtosh} to subtract the background and to compute the track to shower ratio, we obtain:
$$
r_{\text{astro}}^{\text{HESE only}} = 0.25^{+0.11}_{-0.08}
$$
The likelihood is reported in the right panel of Fig.\ref{fig:trtosh} using a yellow curve, \revise{showing also the $1\sigma$ region as a shaded yellow region}. We discuss in the next section why this result is different compared to the one obtained using the throughgoing muon flux.
A summary of the results is reported in Tab.\ref{tab:res}.



\section{Discussion}
\label{sec:disc}
\subsection{An indication of neutron decay ?}
The throughgoing muon flux is based on 36 tracks detected above 200 TeV after 8 years of exposure \cite{Aartsen:2017mau}. This dataset is free from atmospheric muons and negligibly contaminated by atmospheric neutrinos. It may be slightly contaminated by prompt neutrinos but it is for sure dominated by an astrophysical signal. On the other hand the 19 HESE tracks, detected after 7.5 tracks, are expected to be contaminated at level of $\sim 30\%$ by atmospheric muons and atmospheric neutrinos. Moreover the statistic of HESE tracks is a factor 2 smaller that the statistic of throughgoing muons. For these reasons the analysis of the astrophysical track to shower ratio, performed using the throughgoing muon flux, is plausibly more accurate compared to the one performed using HESE only. 


Although the pion decay has been always considered the best mechanism for the production of high energy neutrinos, the neutron decay hypothesis to explain TeV-PeV astrophysical neutrinos is plausible and it was already discussed in literature in \cite{Anchordoqui:2014pca}. \mrevise{However this paper admits that the model should be fine tuned, in order to reproduce the observe data. This is simply related to the processes involved; indeed neutrinos produced by pion decay take about 1/20 of the primary proton's energy, while neutrinos produced by neutron decay would take about $1/1000$ of the neutron energy. Therefore there is a factor $\sim 50$ of difference between the energy budget available for neutrinos from pion decay versus neutrinos from neutron decay. A mechanism able to suppress the photopion production, inside the source, would be required to suppress the neutrino flux expected from the conventional pion decay. This goal may be reached with a particular choice of the target photon field inside the source (for example choosing a peculiar temperature). Another}
possible criticism to the neutron decay scenario would be the over production of events due to the Glashow resonance \cite{Glashow:1960zz} due to the fact that the flux at Earth were dominated by $\bar{\nu}_e$ in this scenario \cite{Barger:2014iua,Biehl:2016psj}. On the other hand in \cite{Palladino:2015uoa} it has been shown that the spectral index and the energy cutoff play a role more important than the production mechanism in the evaluation of the expected number of resonant events. In fact, even assuming a neutron decay scenario, an energy cutoff below 6.3 PeV would nullify the possibility to observe resonant events. 

We also cross checked our procedure computing the number of expected astrophysical showers (given the neutron decay as production mechanism) and comparing it with the number of astrophysical showers resulting after the background subtraction. Since assuming the neutron decay scenario the flavor composition at Earth would be roughly $(\xi_e:\xi_\mu:\xi_\tau)=(2:1:1)$, the expected number of astrophysical HESE showers can be evaluated as follows:
$$
N_s^{\text{astro}} = 4 \pi \mbox{T} \  \int_0^\infty \phi(E) [2 A_e^{\text{eff}}+(1-\eta) A_\mu^{\text{eff}} +A_\tau^{\text{eff}}] \ dE
$$
obtaining $N_s^{\text{astro}}=69.3$ after 7.5 years of exposure. This result is in very good agreement with the $\sim 73$ astrophysical showers found using the background subtraction (see Eq.\ref{eq:nshower}), that is a completely independent method.

In addition to, we notice that this track to shower ratio is also compatible with the neutrino decay scenario \cite{Beacom:2002vi}, assuming normal hierarchy. This scenario has been already investigated in the past. \cite{Pagliaroli:2015rca,Bustamante:2016ciw,Denton:2018aml}.

As a last remark, we notice that the normalization of the throughgoing muon flux $\phi^{100}_\mu$ at 100 TeV is $1.01^{+0.26}_{-0.23}$ (see Sec.4 of \cite{Aartsen:2017mau}) in the usual units of $10^{-18} \rm \ GeV^{-1} \ cm^{-2} \ s^{-1} \ sr^{-1}$, while the normalization of the all flavor HESE flux $\phi^{100}_{3f}$ at the same energy and in the same units is $6.7^{+1.1}_{-1.2}$ \cite{Aartsen:2015knd}. \revise{Let us remark that the normalization of the throughgoing muon flux does not require any assumption on the flavor composition, since this analysis is only sensible to muon neutrinos. On the other hand the normalization of the HESE flux requires an assumption on the production mechanism. Therefore the analysis that follows should be taken as a check of consistency, not as a conclusive result.}
 The ratio between $\phi^{100}_\mu$ and $\phi^{100}_{3f}$ at 100 TeV is therefore equal to:
$$
\frac{\phi^{100}_\mu}{\phi^{100}_{3f}} = 0.15 \pm 0.05
$$
This can be compared to theoretical flavor fraction of muon neutrinos expected in the case of neutron decay, obtaining:
$$
\xi_\mu = P_{e\mu} = 0.225 \pm  0.03
$$
Also this rough estimation, based only on the flux at 100 TeV, supports the neutron decay hypothesis. 

\begin{table}[t]
\begin{center}
\caption{Summary of the results. The expected track to shower ratio for each production mechanism is reported and compared with the astrophysical track to shower ratio obtained using HESE + throughgoing muons and HESE only. The tension between observations and expectations is quoted in terms of number of sigma.}
\label{tab:res}
\begin{tabular}{cccc}
\hline
 & $\pi$ & $\mu$ & $n$ \\
 \hline
$r_{\text{th}} $& $0.21 \pm 0.01 $ & $0.29 \pm 0.04$ &  $0.11 \pm 0.02$  \\
$r_{\text{obs}}^{\text{HESE+TGM}}$  & 2.0$\sigma$ & 2.6$\sigma$ & \revise{$< 1\sigma$} \\
$r_{\text{obs}}^{\text{HESE only}}$  &  \revise{$< 1\sigma$} &  \revise{$< 1\sigma$} & $1.7\sigma$ \\
\hline
\end{tabular}
\end{center}
\end{table}%
\normalsize

\subsection{Alternative interpretation: a complex spectral shape}
\begin{figure}[t]
\centering
\includegraphics[scale=0.45]{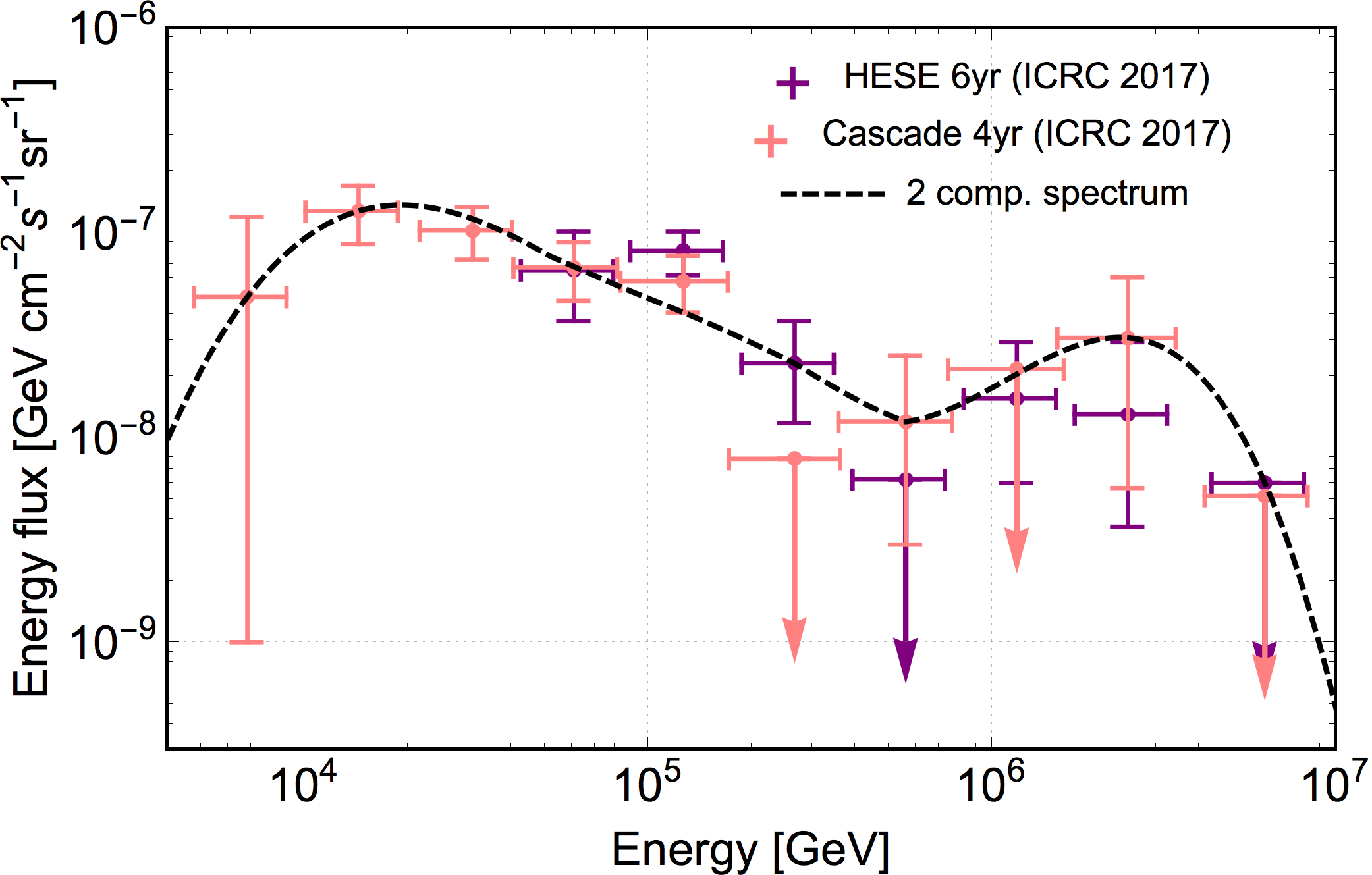}
\caption{Example of non trivial astrophysical neutrino energy spectrum characterized by two peaks. In the same figure 6 years of HESE and 4 years of cascades are reported according to \cite{Aartsen:2017mau}.}
\label{fig:2bump}
\end{figure}
Excluding pessimistic hypotheses, such as a large misidentification of tracks in showers and/or the atmospheric background not under control, 
the result presented above can be also interpreted in a different way, that we are going to discuss in this section. Summarizing, we have seen that the number of HESE tracks obtained from our baseline model is $N_t^{\text{astro}}= 9.3^{+2.6}_{-2.3}$ in 7.5 years (see Eq.\ref{eq:nastro}). Considering only the energy range above 60 TeV, the expectation becomes roughly 80\% of the previous number (see Tab.4 of \cite{Aartsen:2014gkd}). On the other hand 19 HESE tracks have been detected and 6 of them are expected to be background events, resulting in 13 astrophysical tracks. Therefore this number is \revise{approximately} a factor 2 larger than the number expected from the baseline model, that is based on the throughgoing muon flux. 

This discrepancy may suggest a spectrum much more complex than a simple power law flux. 
Let us assume, as an example, that the true astrophysical flux looks like the toy spectrum represented in Fig.\ref{fig:2bump} with a dashed black curve. Under the hypothesis of pion decay, this flux would give rise to $\sim 10$ HESE tracks above 60 TeV. Considering the background, we would obtain a total of 16 HESE tracks expected versus 19 observed, with a non significant tension accounting for the Poissonian uncertainty. The spectrum of Fig.\ref{fig:2bump} would suggest a flux above 200 TeV harder than $E^{-2}$. In \cite{Aartsen:2015zva} we found indication for a hard throughgoing muon spectrum, characterized by $E^{-\alpha}$ with $\alpha=1.91 \pm 0.20$. Nowadays, with the increasing of the exposure, the data seems to prefer  a softer spectrum, characterized by $\alpha \simeq 2.2 \pm 0.1$, as reported in \cite{Aartsen:2017mau}. On the other hand the hypothesis of a more complex spectral shape is worthy of being investigated, since a power law neutrino spectrum is only expected from sources in which neutrinos are produced via proton-proton interaction \cite{Kelner:2006tc}, while it is not compatible with neutrinos produced in sources dominated by $p\gamma$ interaction \cite{Kelner:2008ke}. For example the toy spectrum represented in Fig.\ref{fig:2bump} may be produced by two different populations of sources dominated by $p\gamma$ interaction.

\revise{For the sake of completeness, we need to remark that all the paper is based on the assumption that the flavor composition is energy independent between roughly 10 TeV and 10 PeV. In environments with strong magnetic fields, the flavor composition may be energy dependent going from the pion decay scenario to the damped muon scenario with the increasing of the neutrino energy. However this scenario goes in the opposite direction compared to our findings, therefore it cannot be used as a possible explanation for our results.}

\section{Conclusion}
\label{sec:conc}
In this work we investigate the track to shower ratio suggested by astrophysical neutrinos after 8 years of observations in IceCube. We compare it with the ones expected from three theoretical scenarios, namely the pion decay, the damped muons and the neutron decay. We use the natural parametrization to compute the oscillations of astrophysical neutrinos and we take advantage of the most recent IceCube measurements, by using a broken power law spectrum that is in agreement with all the data between $\sim 30$ TeV and few PeV. Moreover we used the flux of throughgoing muons to evaluate the expected number of astrophysical HESE tracks and we take into account that background that can affect HESE showers.
We conclude that the observed track to shower ratio is fully consistent with the neutron decay scenario while it is in tension at level of $2\sigma$ and $2.6\sigma$ with the standard pion decay scenario and with the damped muons one, respectively. This result differs from \cite{Mena:2014sja}, in which a null track to shower ratio was favored using all HESE, although also in that case the neutron decay scenario was the best option among the standard astrophysical mechanisms. It is also different compared to \cite{Aartsen:2015knd}, in which only events above 60 TeV are considered and the damped muon scenario is the best candidate mechanism. In addition to our use of the most updated datasets, the main difference is that our work does not rely on the background that affects astrophysical HESE tracks, that represents the biggest source of uncertainties in the computation of the track to shower ratio. To tackle this problem the number of expected astrophysical HESE tracks is computed thanks to the well measured throughgoing muon flux, showing that the extrapolation below 200 TeV plays only a minor role. In principle all these three methods should give the same results; these differences may stem from the uncertainties of the poorly known atmospheric muon background. 

Another possibility is that the spectrum of astrophysical neutrinos is much more complex than a power law flux. We have shown that an energy spectrum with two peaks may alleviate the tension between HESE and throughgoing muons, partially recovering the compatibility with the pion decay scenario. 

Both the previous possibilities are worthy of being investigated. If the indication for a neutron decay scenario were confirmed and improved in the future, it would have an impact on the models that aim to explain the high energy neutrino emission, given the fact that in most of the models neutrinos are expected to be produced by pion decays and not by neutron decay, although the last possibility has been already considered in the scientific literature. On the other hand if the spectrum is much more complex than a power law flux, this may also have an impact on several aspects related to the interpretation of astrophysical neutrinos and to the multi-messenger connection with the diffuse flux of $\gamma$-rays.

\vspace{0.1 cm}
{\bf Acknowledgments.} This project has received funding from the European Research Council (ERC) under the European Union’s Horizon 2020 research and innovation programme (Grant No. 646623). I thank Markus Ackermann, Shan Gao, Francesco Vissani and Walter Winter for the useful discussions.

\bibliographystyle{ieeetr}
\bibliography{bibliography.bib}
 
\end{document}